This is the Accepted Manuscript of this published paper: Vacher, L. G., Phan, V. T. H., Bonal, L., Iskakova, M., Poch, O., Beck, P., Quirico, E., Ogliore, R. C. (2025). Amides from the carbonaceous asteroid (162173) Ryugu: Nanoscale spectral and isotopic characterizations. Meteoritics & Planetary Science. https://doi.org/10.1111/maps.70019

# Amides from the carbonaceous asteroid (162173) Ryugu: nanoscale spectral and isotopic characterizations

*L. G. Vacher[1], V. T. H. Phan[1], L. Bonal[1], M. Iskakova[2], O. Poch[1], P. Beck[1], E. Quirico[1], & R. C. Ogliore[2]

*Corresponding author: lionel.vacher@univ-grenoble-alpes.fr

[1]Univ. Grenoble Alpes, CNRS, IPAG, 38000 Grenoble, France

[2]Department of Physics, Washington University in St. Louis, St. Louis, MO, USA.

**Abstract**

C-type asteroids, such as asteroid (162173) Ryugu, may have played a key role in delivering light elements to early Earth. Nitrogen (N)-bearing molecules have been chemically identified in some Ryugu grains, and based on the faint 3.06 µm absorption band observed by the hyperspectral microscope MicrOmega, NH-bearing compounds seem to be spread at the global scale in the collection. However, the chemical forms of these NH-bearing compounds —whether organic molecules, ammonium ($NH_4^+$) salts, $NH_4^+$- or NH-organics-bearing phyllosilicates, or other forms— remain to be better understood. In this study, we report the characterization of two Ryugu particles (C0050 and C0052) using infrared spectroscopy at millimeter, micrometer, and nanometer scales, along with NanoSIMS techniques to constrain the nature and origin of NH-bearing components in the Ryugu asteroid. Our findings show that Ryugu's C0052 particle contains rare (~1 vol.%), micrometer-sized NH-rich organic compounds with peaks at 1660 $cm^{-1}$ (mainly due to C=O stretching of the amide I band) and 1550 $cm^{-1}$ (mainly due to N-H bending vibration mode of the amide II band), indicative of amide-related compounds. In contrast, these compounds are absent in C0050. Notably, N isotopic analysis reveals that these amides in C0052 are depleted in $^{15}$N ($\delta^{15}N \simeq -200$ ‰), confirming their indigenous origin, while carbon (C) and hydrogen (H) isotopic compositions are indistinguishable from terrestrial values within errors. The amides detected in C0052 could have formed through hydrothermal alteration from carboxylic acids and amines precursors on the Ryugu's parent planetesimal. Alternatively, they could have originated from the irradiation of $^{15}$N-depleted N-bearing ice by ultraviolet light or galactic cosmic rays, either at the surface of the asteroid in the outer Solar System or on mantle of interstellar dust grains in the interstellar medium. Amides delivered to early Earth by primitive small bodies such as asteroid Ryugu may have contributed to the prebiotic chemistry.

## 1. Introduction

In December 2019, the Japan Aerospace Exploration Agency (JAXA) Hayabusa2 spacecraft returned to Earth samples from the surface (Chamber A) and possibly from the sub-surface (Chamber C) of the near-Earth carbonaceous (C-type) asteroid (162173) Ryugu (Watanabe et al., 2019; Tachibana et al., 2022). Preliminary laboratory analyses revealed that these samples are mineralogically, chemically, and isotopically similar to rare volatile-rich Ivuna-type carbonaceous (CI) chondrites (Hopp et al., 2022; Yada et al., 2022), whose average compositions are similar to that of the solar photosphere (Palme et al., 2014). The returned samples provide strong evidence of significant aqueous alteration, as seen by the widespread occurrence of secondary minerals, including phyllosilicates (serpentine and saponite), dolomite, magnetite, and pyrrhotite. These minerals likely formed through low-temperature (<100°C) fluid circulation on the parent asteroid (T. Nakamura et al., 2022; Yada et al., 2022; Ito et al., 2022). Organic matter (OM), accounting for ~3 wt.% of the total mass of the studied particles (Yabuta et al., 2023), shares characteristics with OM found in other carbonaceous chondrites (Dartois et al., 2023; Naraoka et al., 2023; Bonal et al., 2024; Quirico et al., 2024). The OM is dispersed throughout the matrix, primarily as diffuse intergranular matter associated with phyllosilicates and nanoglobules (Yabuta et al., 2023; Mathurin et al., 2024; Phan et al., 2024; Gainsforth et al., 2024; Stroud et al., 2024), suggesting that OM synthesis and preservation occurred under low-temperature aqueous conditions within the Ryugu's parent planetesimal (Potiszil et al., 2023).

In JAXA's curation facility (CF), within clean and $N_2$-purged chambers, spectroscopy measurements of Ryugu samples were performed at millimeter scale using a commercial Fourier Transform InfraRed spectrometer (VIR-300, JASCO Corporation; named "FTIR-CF" in the following) (Yada et al. 2022; Hatakeda et al., 2023) and at sub-millimeter scale using the MicrOmega hyperspectral near-infrared microscope (Pilorget et al., 2022, 2024). These analyses revealed a faint ~3.06-μm band in the bulk samples from Chambers A and B (Yada et al., 2022), and strong absorption bands at 3.06 μm and 3.24 μm on a few particles (Pilorget et al., 2022). These absorption bands are attributed to N-H bond vibrational modes in NH-rich compounds. These compounds could be $NH_4^+$ phyllosilicates, $NH_4^+$ hydrated salts, $NH_4^+$ phosphorous-rich grains (Pilorget et al., 2024), and/or NH-bearing organic molecules (e.g., amines or amides) possibly trapped within the interlayer spaces of phyllosilicates in Ryugu samples (Viennet et al., 2023). Since this first compositional analysis via infrared (IR) spectroscopy, NH-bearing organic molecules have been firmly detected in Ryugu samples via analytical chemistry (Yoshimura et al., 2023; Takano et al., 2024). However, whether these organic molecules, along with other NH-bearing compounds, contribute to the global ~3.06-μm band detected in Ryugu remains to be investigated. Identifying the nature and distribution of NH-bearing phases in Ryugu samples is crucial for constraining the N reservoirs that made up primitive small bodies (Füri and Marty, 2015; Hily-Blant et al., 2019; Grewal, 2022). For instance, $NH_4^+$ has been detected at the surfaces of the dwarf planet Ceres, mainly associated to phyllosilicates (King et al., 1992; De Sanctis et al., 2015), and also on the surface and dust grains of comet 67P/Churyumov-Gerasimenko, in the form of $NH_4^+$ salts (Poch et al., 2020; Altwegg et al., 2020, 2022).

While conventional IR methods can analyze the functional groups of C and N in Ryugu particles down to a size of ~25 × 25 μm$^2$ using a GLOBAR thermal light source, novel IR techniques with nanoscale spatial resolution, such as atomic force microscope-based infrared spectroscopy (AFM-IR), can overcome the diffraction limit and identify mineral or organic phases at the nanometer scale (Mathurin et al., 2019; Kebukawa et al., 2019b; Phan et al., 2022, 2024). Coordinated analyses of AFM-IR and nanoscale secondary ion mass spectrometry (NanoSIMS) offer a new powerful approach (Kebukawa et al., 2023), enabling the identification and isotopic characterization of NH-bearing compounds in Ryugu grains at similar spatial resolutions to constrain their origins and evolutionary history.

In this study, with the aim to search for NH-bearing components in Ryugu particles, we report bulk and in situ characterizations of several fragments extracted from two Ryugu particles (C0050 and C0052) using multi-scale IR techniques, including reflectance spectroscopy at the millimeter scale, transmission spectroscopy at the micrometer scale, and atomic force microscope based infrared spectroscopy (AFM-IR) at the nanometer scale. The latter technique enabled the identification of amides in those Ryugu fragments. The combination of AFM-IR and NanoSIMS microscale techniques enabled to investigate the C, N, and H isotopic composition of these amides.

## 2. Methods

### 2.1. Selection of Hayabusa2 particles and sample preparation

The selection process of the Ruygu particles available for Announcement of Opportunity (AO) allocation was made using the FTIR spectra available in the Ryugu Sample Database in January 2022, as detailed in Supplementary material S1. Briefly, to avoid spectral ambiguities related to the fact that carbonate and $NH_4^+$ can have spectral absorptions at similar location around 3.4 µm and 7 µm (Lin-Vien et al., 1991; Petit et al., 2006), we have selected five particles whose FTIR spectra show no carbonate absorption band from about 3.8 µm to 4.0 µm (Figure S1). The JAXA curation team allocated to our consortium the particles C0050 (2.2 mg) and C0052 (2.1 mg), which were then sent to the Institut de Planétologie et d'Astrophysique de Grenoble (IPAG, France), arriving in September 2022.

To prevent contamination, the Hayabusa2 sealed containers were opened in a controlled environment, using a glove bag flushed with $N_2$ (RH ≤ 5 % and $O_2$ < 0.1 %). The bulk particles were then transferred to a vacuum sapphire viewport for reflectance analysis. After the initial reflectance analysis, the particles were gently crushed in the glove bag. Small fragments were selected, crushed between diamond windows, and then placed in an environmental cell for transmission µ-FTIR analysis. Following transmission analysis, the environmental cell was opened, and the diamond windows were transferred into the AFM-IR at IPAG to detect IR vibrational signatures at the nanometer scale. After AFM-IR investigations, the diamond windows were coated with gold and subjected to Scanning Electron Microscope (SEM) investigations (Figure S2). After SEM observations, the coated diamond windows were then mounted on an in-house NanoSIMS holders for isotopic characterizations at the Laboratory for Space Sciences at Washington University (Wash U) in St. Louis (MO, USA).

### 2.2. Reflectance Spectroscopy (SHADOWS)

Reflectance spectra of particles C0050 and C0052 were acquired between 2 µm to 4.2 µm using the spectro-gonio radiometer SHADOWS (Spectrophotometer with Hanging Angles for the Detection of Weak Signals, Potin et al. 2018) at IPAG. This instrument is specially designed to measure dark samples with weak reflectance spectra from the visible to the near-infrared (350–5000 nm). The samples were illuminated with a beam ≤1.7 mm in diameter, at normal incidence (0°), and with an emergence angle of 30° (Potin et al., 2018). The spectrum was sampled every 20 nm, and the spectral resolution was 20 nm from 1.60 µm to 2.82 µm, and 41 nm from 2.84 µm to 4.20 µm. Surfaces of Spectralon and Infragold (LabSphere Inc.) were measured, from 0.38 µm to 2.00 µm and from 1.20 µm to 4.8 µm, respectively, under the same conditions as the samples and used as references to calibrate the signal measured on the samples and obtain absolute values of reflectance factors, as described in Potin et al. (2018).

### 2.3. Micro-FTIR Spectroscopy (μ-FTIR)

The IR spectra were collected using a Bruker Hyperion 3000 IR microscope equipped with a liquid $N_2$ cooled mercury–cadmium–telluride (MCT) detector at IPAG. A 15× objective lens was used to focus the IR beam onto the samples, with a maximum spot size of 100 × 100 μm². Spectral resolution was set at 4 cm$^{-1}$ across the range of 4000–650 cm$^{-1}$. Fragments from both particles of typically ~50−100 μm in size were manually selected inside a $N_2$-flushed glove bag using an Olympus SZX16 stereomicroscope, then were transferred onto a 3 × 0.5 mm diamond window and crushed with another window using a custom press designed at IPAG. These windows were either placed in a custom environmental chamber to maintain a controlled atmosphere or analyzed directly in air. Spectral data were processed using in-house MATLAB code or Igor Wavemetrics, with baseline corrections applied via the "Baseline Fit" Matlab function. Normalization was performed by setting the absorbance of the Si-O peak (~1000 cm$^{-1}$) to 1.

### 2.4. Atomic Force Microscope-Based Infrared Spectroscopy (AFM-IR)

AFM-IR analyses were conducted on Ryugu samples following μ-FTIR measurements in a dry compressed air environment. Nanoscale IR imaging and absorption spectra were collected using a nanoIR3s system (Bruker) at IPAG. The system's AFM probe generated topographical images and measured IR absorption via photo-thermal expansion. The excitation laser source, a Carmina laser (APE GmbH, Germany) covered the mid-IR range (2000−700 cm$^{-1}$), while the Firefly IR laser (M Squared, UK) provide light in the 2700−4000 cm$^{-1}$.

Two modes, contact mode and tapping IR mode, were used for spectroscopy and imaging. Laser intensity and alignment were optimized before data collection. For AFM-IR imaging, the APE laser was used in tapping IR mode with a laser power ranging of 25−40 % and a pulse rate of 340−400 kHz, conditions optimized to prevent damage to macromolecular organics due to laser irradiation. The analytical protocol employed in this study was validated in previous works on meteorites and coals (Phan et al., 2022, 2023, 2024).

AFM-IR images were collected across specific wavenumber ranges, including 2000−700 cm$^{-1}$ for carbonyl (C=O) stretching, sp$^2$ aromatic (C=C), N-H bending mode, methylene ($CH_2$), and/or carbonate and/or $NH_4^+$ bending mode, and silicate (Si-O) stretching. Images at 1660 cm$^{-1}$ and 1550 cm$^{-1}$ were also obtained to explore double peaks in these absorption bands, potentially indicating organic compounds like amide groups (e.g., -C(=O)-NH-). In the 3800−2700 cm$^{-1}$ range, AFM-IR images were captured in contact mode using the Firefly laser targeting absorption bands related to −OH stretching (3600–3700 cm$^{-1}$), asymmetric stretching of $CH_3$ (2960 cm$^{-1}$), asymmetric stretching of $CH_2$ (2930 cm$^{-1}$), symmetric stretching of $CH_3$ (2880 cm$^{-1}$), and symmetric stretching of $CH_2$ (2860 cm$^{-1}$). To visualize the spatial distribution of these components, composite RGB color images were generated by superimposing absorption images using

Analysis Studio and MountainMapSPIP@ software. Images were realigned to correct for thermal drifts with the scan speeds set to 0.1 Hz, and scan sizes ranging from 300 × 300 points to 500 × 500 points depending on the Region of Interest (ROI).

Single-point AFM-IR spectra were obtained for Ryugu grains, C0052 and C0050 using contact mode with both the APE laser (2000−700 $cm^{-1}$) and the Firefly laser (3800−2700 $cm^{-1}$) for comprehensive analysis. To ensure high-quality spectra and preserve sample integrity, low laser power (1.22−5.03%) and a pulse rate of 240−300 kHz were maintained (Phan et al., 2022, 2023, 2024). Spectra were optimized with constant laser power, detected via an IR-sensitive photodetector with a wavenumber spacing of 4 $cm^{-1}$, co-averages of three spectra and five spectra for APE and Firefly lasers. For the 2000−700 $cm^{-1}$ range, a gold-coated semi-tap probe (PR-EX-TnIR-A-10) was used to reduce artifacts from silicon IR absorption, allowing operation in both contact and tapping IR modes. In the 4000–2700 $cm^{-1}$ range, contact mode was employed with the “pure contact” model (PR-EX-nIR2-1).

## 2.5. NanoSIMS

Carbon, N and H isotopic measurements of C0052 particle were performed with the Cameca NanoSIMS 50 ion microprobe instrument at Wash U in St. Louis. The samples were measured in two sessions using a ~2.7−3.5 pA $Cs^+$ primary beam focused to 100 nm. In the first session, we collected $^{12}C^-$, $^{13}C^-$, $^{12}C^{14}N^-$, $^{12}C^{15}N^-$ and $^{28}Si^-$ for C and N isotopes. In the second session, we collected $H^-$ and $D^-$ for H isotopes. Before measurement, the analyzed areas were pre-sputtered on an 8 × 8 µm² area using a higher primary beam current, with the duration adjusted until stable secondary ion signals were achieved. Then, the primary beam was rastered over a 5 × 5 µm area, divided into 256 × 256 pixels. For C and N isotopes, the mass resolving power (MRP) was set to ~5700 to resolve $^{12}CH^-$ and $^{13}C^{14}N^-$ interferences on $^{13}C$ and $^{12}C^{15}N^-$, respectively. Optimum count rates for $^{12}C^-$ and $^{12}C^{14}N^-$ were obtained by applying a compromise EOS value that maximized the count rates for both species. For H isotopes, the MRP was set to ~1700. We used a kerogen standard (pressed into copper stub) from Chert (Warrawoona Group, Australia, 3.5 billion years) containing 64 wt.% of C, with an wt.% C/N ratio of 181.59, wt.% H/C ratio of 0.3, $\delta^{13}C_{PDB}$ of −34 ‰, $\delta^{15}N_{AIR}$ of 6 ‰ (Beaumont and Robert, 1999), and $\delta D_{SMOW}$ = −105 ‰ as reference material to correct for instrumental mass fractionation (IMF). The IMF was measured to be −111 ± 16 ‰ (2σ) for C, −163 ± 24 ‰ (2σ) for N, and −120 ± 216 ‰ (2σ) for H (Figure S5 and Tables S3−S4). We collected a total of 30 cycles and manually removed cycles until counts are stable. We typically achieved a counting statistic of ≤ 4 ‰ for $\delta^{13}C$ (2σ), ≤ 32 ‰ (2σ) for $\delta^{15}N$ and ≤ 32 ‰ (2σ) for $\delta D$ on kerogen. The NanoSIMS image data were processed using an in-house MATLAB code and corrected for stage or beam drift. Delta values were further corrected for a 44 ns dead time and for quasi-simultaneous arrival (QSA) using the method described by Ogliore et al. (2021), which allows multiplicative effects to cancel out when the standard and unknowns have similar count rates. However, the secondary count rates of the kerogen standard are approximately two orders of magnitude higher than those of the unknowns for $^{12}C^-$ and $^1H^-$ (Tables S3–S4), resulting in estimated QSA corrections of ~25 ‰ for $\delta^{13}C$ and ~15 ‰ for δD. These corrections, however, remain within the uncertainties of the final measurements of the unknown samples (≥ 28 ‰ for $\delta^{13}C$ and ≥ 245 ‰ for δD at 2σ, Table 1) and therefore, were not applied. Pixels from the $\delta^{13}C$, $\delta^{15}N$, and $\delta D$ NanoSIMS images were resized so that the pixel size was larger than the beam size. Carbon, N and H isotopic ratios ($^{13}C/^{12}C$, $^{15}N/^{14}N$ and D/H) are reported as delta values, relative to PDB ($\delta^{13}C_{PDB}$ = $(^{13}C/^{12}C)_{sample}/(^{13}C/^{12}C)_{PDB}$ – 1] × 1000, PDB = Pee Dee Belemnite), AIR ($\delta^{15}N_{AIR}$ = $(^{15}N/^{14}N)_{sample}/(^{15}N/^{14}N)_{AIR}$ – 1] × 1000), and SMOW ($\delta D_{SMOW}$ = $[(D/H)_{sample}/(D/H)_{SMOW}$ – 1] × 1000, SMOW = Standard Mean Ocean Water), respectively.

## 3. Results

### 3.1. Reflectance spectroscopy

The IR reflectance spectra of particles C0050 and C0052, acquired under $N_2$ in the wavelength range from 2 µm to 4 µm using the spectro-gonio radiometer SHADOWS, are shown in Figure 1. These spectra can be compared to the MicrOmega spectra acquired inside the curation facility[1], the spectrum of Ryugu bulk samples from Chamber C (Yada et al., 2022), and the CI chondrite Orgueil (Potin et al., 2020). The spectra of particles C0050 and C0052 exhibit a prominent absorption band at ~2.7 µm, due to hydroxyl (−OH) in phyllosilicates (T. Nakamura et al., 2022). This observation is consistent with the FTIR-CF spectra obtained from the bulk Ryugu samples from Chamber C (Yada et al., 2022), the CI Orgueil and other hydrated carbonaceous chondrites (Potin et al., 2020; Amano et al., 2023). However, our reflectance spectra show a broader absorption feature between ~2.7 µm to 3 µm compared to the MicrOmega spectra acquired in JAXA's curation facility, suggesting a slight absorption of molecular water despite our precautions during sample handling and storage. No additional spectral features, such as those associated with organic molecules (3.4 µm and 3.5 µm) or carbonate (3.4 µm and 3.9 µm), are observed on the SHADOWS spectra.

Notably, the MicrOmega and SHADOWS spectra of particles C0050 and C0052 shown in Figure 1 do not have absorption feature around 3.06 µm and 3.24 µm that could be attributed to $NH/NH_4^+$-bearing compounds, although the FTIR spectra on which we based our grain selection show absorptions around these wavelengths (Figure S1). Since both FTIR-CF and MicrOmega instruments operate within JAXA's curation facility, the differences between their spectra of the same grains are either due to instrumental artefact(s) of the FTIR-CF measurement (possibly due to a water ice film on the detector, as mentioned in

---

[1]The MicrOmega spectra were available from the Ryugu Sample Data System after the first AO allocation deadline. Therefore, our grain selection was only based on the FTIR spectra.

Hatakeda et al. 2023) and/or to a change of the grain composition between the FTIR-CF and MicrOmega measurements (potentially due to a loss of volatile compounds due to the vacuum tweezer used to manipulate the grains within the facility, and/or due to heating of these extremely dark grains during their illumination). Since the spectra measured outside the curation facility resemble the MicrOmega spectra, we favor the first hypothesis, although the second one might not be totally excluded.

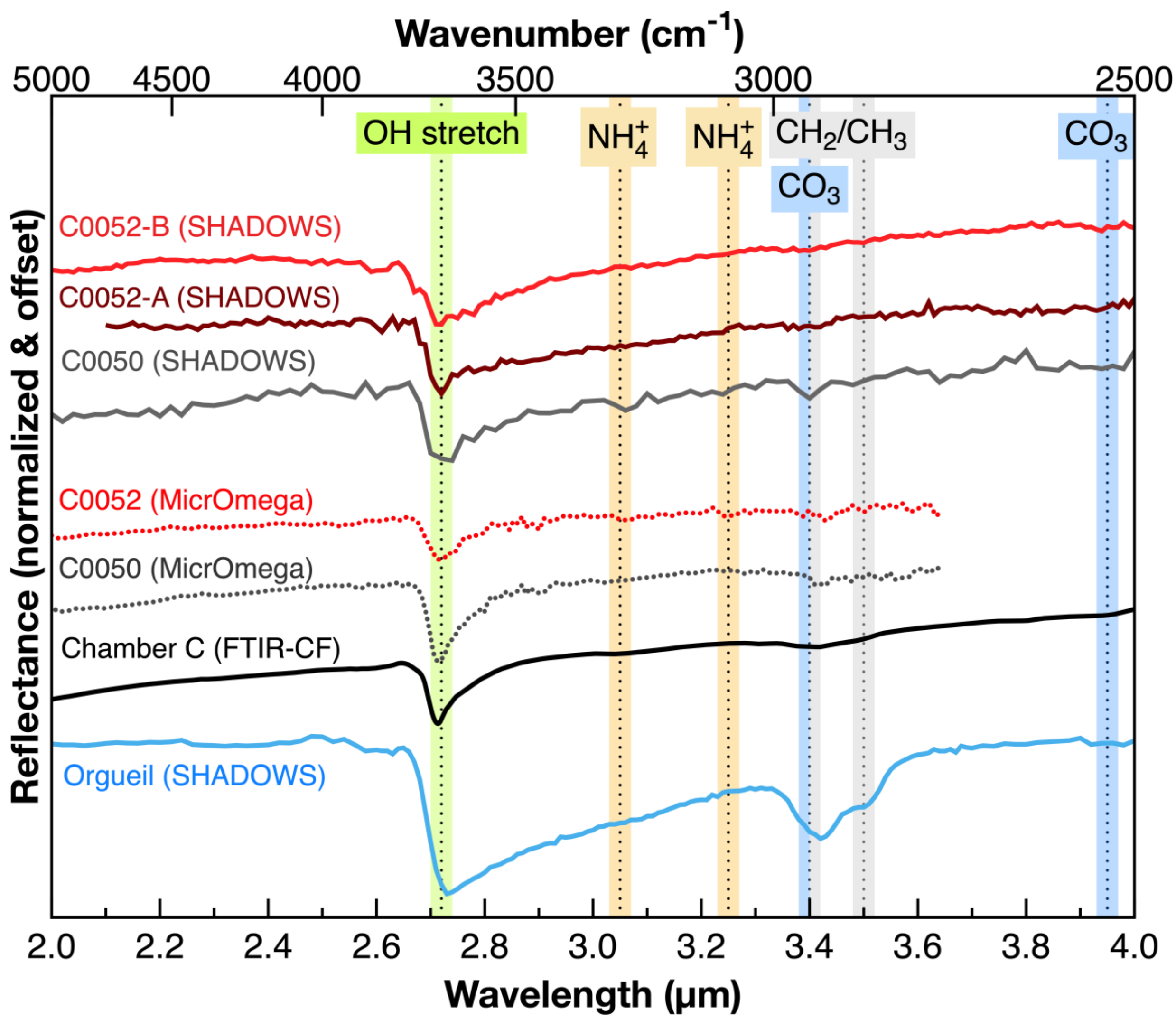

**Figure 1.** Reflectance spectra between 2 µm to 4 µm (normalized to 2.5 µm) of Ryugu particles C0052 (red) and C0050 (grey) using the spectro-gonio radiometer SHADOWS, compared to MicrOmega measurements (dotted spectra), averaged Ryugu samples from Chamber C collected in the JAXA's curation facility (FTIR-CF) (solid black spectrum, Yada et al. 2022), and a powder of the CI chondrite Orgueil measured under vacuum at room temperature using the SHADOWS instrument (solid blue spectrum, Potin and Beck 2019). The particle C0052 has been measured on both sides up and down (C0052-A and B). Typical OH (2.7 µm, in green), $NH_4^+$ (3.05 µm and 3.25 µm, in yellow), organic molecules (3.4 µm and 3.5 µm, in grey), and carbonate (3.4 µm and 3.9 µm, in blue) band positions are highlighted by vertical black dotted lines.

### 3.2. Transmission spectroscopy

The μ-FTIR spectra of crushed fragments from particles C0050 and C0052, obtained under $N_2$ and atmospheric conditions, are depicted in Figure 2. The spectra acquired under $N_2$ display homogeneity, with consistent peak positions and shapes indicating the presence of various compounds, such as Mg-rich phyllosilicates (notably −OH stretching at ~3690 $cm^{-1}$ and Si-O stretching at ~1000 $cm^{-1}$), organic compounds (including $CH_2$ and $CH_3$ stretching modes at ~2950−2860 $cm^{-1}$, and double bond C=C at ~1600 $cm^{-1}$ for C0050, respectively), and possibly minor contribution from carbonates ($CO_3$ bending et in-plane modes at ~1450 $cm^{-1}$ and 880 $cm^{-1}$). These IR signatures are similar to those previously measured in other Ryugu particles (Dartois et al., 2023; Mathurin et al., 2024; Phan et al., 2024; Bonal et al., 2024; Yesiltas et al., 2024; Quirico et al., 2024) and CI chondrites, such as Orgueil and Ivuna (Beck et al., 2010).

The IR spectra acquired under atmospheric conditions are similar to those obtained under $N_2$, but exhibit additional features, including a broad O-H stretching band at ~3400 $cm^{-1}$ and a H-O-H bending band at ~1650 $cm^{-1}$. These bands are less prominent in samples analyzed under $N_2$ environment and are attributed to water molecules loosely adsorbed to minerals and as interlayer molecular water in phyllosilicates due to terrestrial contamination. In some spectra obtained from particle C0052 (Figure S3), we also observed a weak band near ~3250 $cm^{-1}$ (7-G2-1 and 7-G1-3 from Figures 2 and S3). This spectral signature overlaps with the main asymmetric stretching vibration mode of $NH_4^+$ (~3250 $cm^{-1}$; Fastelli et al. 2020; Petit et al. 2006), possibly suggesting the presence of N-H containing species. However, this band also coincides with the vibration mode of water ice at ~3000−3200 $cm^{-1}$ (depending on the temperature), which may have condensed on the liquid-nitrogen-cooled detector of the FTIR microscope during analysis (Hatakeda et al., 2023). Although we carefully collected and subtracted each background before collecting the spectrum of the sample to minimize the appearance of a water ice signal, we cannot exclude the possibility that this broad band is caused solely by this instrumental artifact.

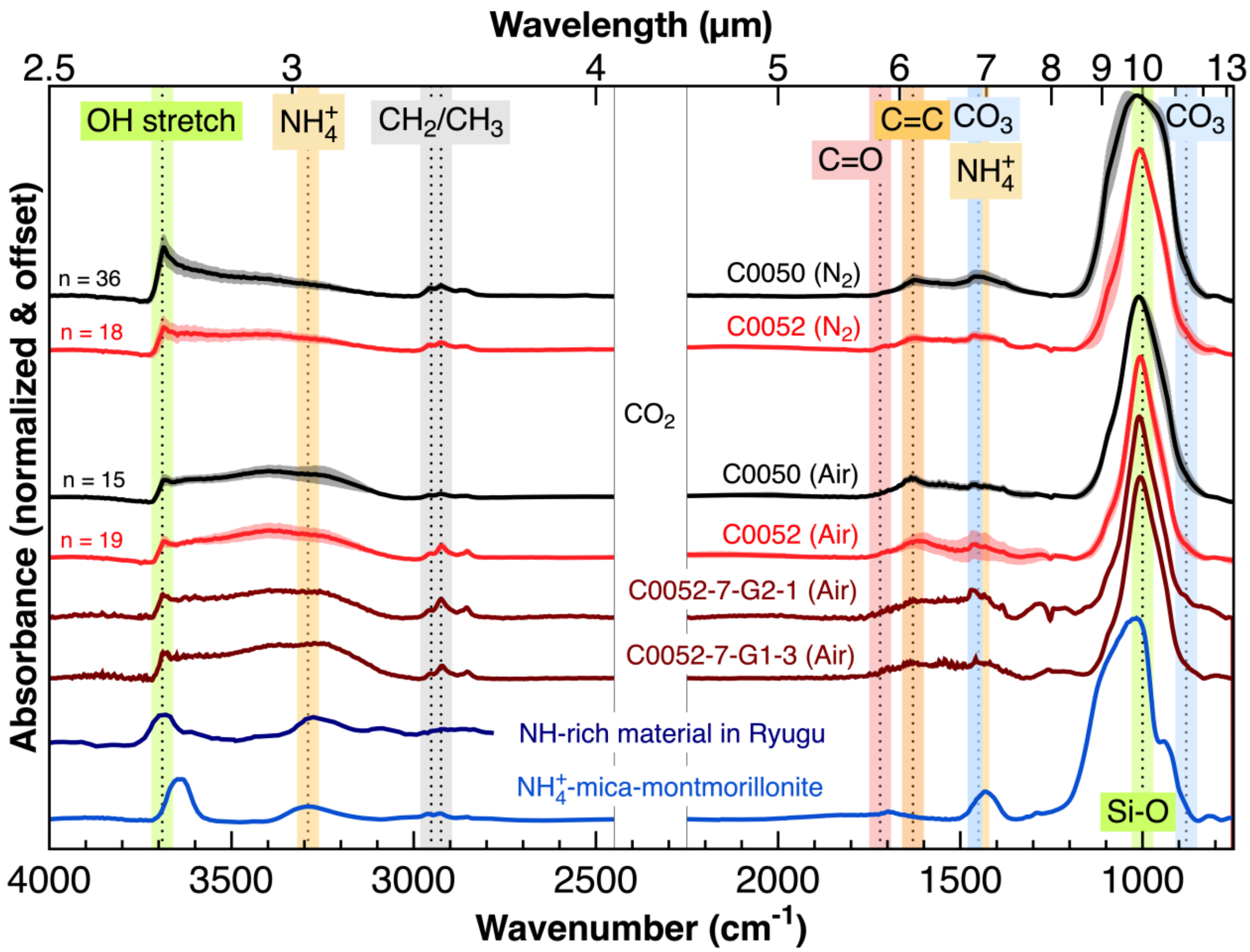

**Figure 2.** μ-FTIR spectra (normalized to 1000 cm$^{-1}$) of multiple crushed grains extracted from particles C0050 and C0052, collected under $N_2$ or air conditions. For particles C0050 (black) and C0052 (red), all spectra represent averages, with the 1σ error shown as shaded areas. The spectra of specific regions in particle C0052 (7-G2-1 and 7-G1-3), shown individually for comparison (dark red), display an additional weak band near ~3250 cm$^{-1}$, possibly indicating the presence of N-H containing species. A similar absorption band attributed to NH-rich material has also been observed on a few Ryugu particles (dark blue, Pilorget et al., 2022). A spectrum of $NH_4^+$-mica-montmorillonite reference material (light blue, analyzed with the same condition for comparison), which displays a similar absorption feature near ~3250 cm$^{-1}$, is also shown for comparison. The peaks at 2360 cm$^{-1}$ corresponding to atmospheric $CO_2$ are hidden for clarity. Colored dashed lines indicate the positions of the following absorption bands: green for phyllosilicate OH (3690 cm$^{-1}$) and Si-O bands (1000 cm$^{-1}$), yellow for $NH_4^+$ bands (3250 cm$^{-1}$ and 1450 cm$^{-1}$), gray for aliphatic C-H bands (2960 cm$^{-1}$ for $CH_3$ asymmetric stretching and 2925 cm$^{-1}$ for $CH_2$ symmetric stretching), red for the carbonyl (C=O) band (1720 cm$^{-1}$), orange for the aromatic (C=C) band with some contribution from water bending modes (~1600 cm$^{-1}$), and blue for carbonate ($CO_3$) bands (1440 cm$^{-1}$ and 880 cm$^{-1}$).

### 3.3. AFM-IR Imaging and Spectroscopy

Using the Tapping IR mode, we investigated the spatial distribution of different functional groups with IR signatures across the entire surface of the C0050 and C0052. A 5 × 5 μm ROI of C0052 was recorded with a higher spatial resolution of 500 × 500 points (obtaining theoretical spatial resolution imaging of ~10 nm) (Figure 3a) (Phan et al., 2022). AFM-IR images obtained at various wavenumbers reveal the distribution of different absorptions: 1660 $cm^{-1}$ (red), 1550 $cm^{-1}$ (yellow), and 1450 $cm^{-1}$ (blue) attributed to $CH_2$ bending or $CO_3^{2-}/NH_4^+$ bending mode, and phyllosilicate (Si-O) at 1000 $cm^{-1}$ (green, Figure 3b−e). Figures 3f and 3g display two composite RGB images, showing IR intensities at 1660 $cm^{-1}$, 1550 $cm^{-1}$, and 1000 $cm^{-1}$ (Figure 3f), and at 1660 $cm^{-1}$, 1450 $cm^{-1}$, and 1000 $cm^{-1}$ (Figure 3g). These maps, along with AFM-IR spectra from 2000−700 $cm^{-1}$, reveal three domains: *(i)* a prominent Si-O stretching band around 1000 $cm^{-1}$ in phyllosilicates (green areas in Figures 3e−g), *(ii)* strong absorption peaks at 1660 $cm^{-1}$ and 1550 $cm^{-1}$, unusual for OM found in meteorites or Ryugu sample (Phan et al., 2022; Phan et al., 2024; Mathurin et al., 2024) and *(iii)* small blue regions (Figure 3d) suggesting either $CH_2$ bending indicative of aliphatic organic compounds, or carbonates.

To gain more detailed insights, we obtained numerous single-point AFM-IR spectra in the 2000−700 $cm^{-1}$ range. The average B spectrum ($n$ = 48) shows a strong 1000 $cm^{-1}$ peak typical of phyllosilicates in Ryugu (Figure 3). The C spectrum shows two distinct yet concurrent peaks at 1450 $cm^{-1}$ and 880 $cm^{-1}$, predominantly associated with the bending and in-plane formation modes characteristic of carbonate minerals, such as dolomite and/or calcite (Phan et al., 2022, 2024). Five spectra from the orange and pink areas (A1−A5) show peaks at 1660 $cm^{-1}$ and 1550 $cm^{-1}$, distinct from the typical OM in Ryugu particles (Phan et al., 2024). The position and relative intensity of these two peaks are commonly attributed to the amide group (e.g., -C(=O)-NH-), with the peak at 1660 $cm^{-1}$ mainly due to C=O stretching (so-called "amide I" band) and 1550 $cm^{-1}$ mainly due to N-H bending vibration mode (so-called "amide II" band). SEM-EDS data indicate that these organic-rich regions are enriched in N in comparison to the surrounding areas (Figure S2). A similar spectral signature was also found in another area in the 7-G1-3 grain, showing double peaks at 1660 $cm^{-1}$ and 1550 $cm^{-1}$ (Figure S4).

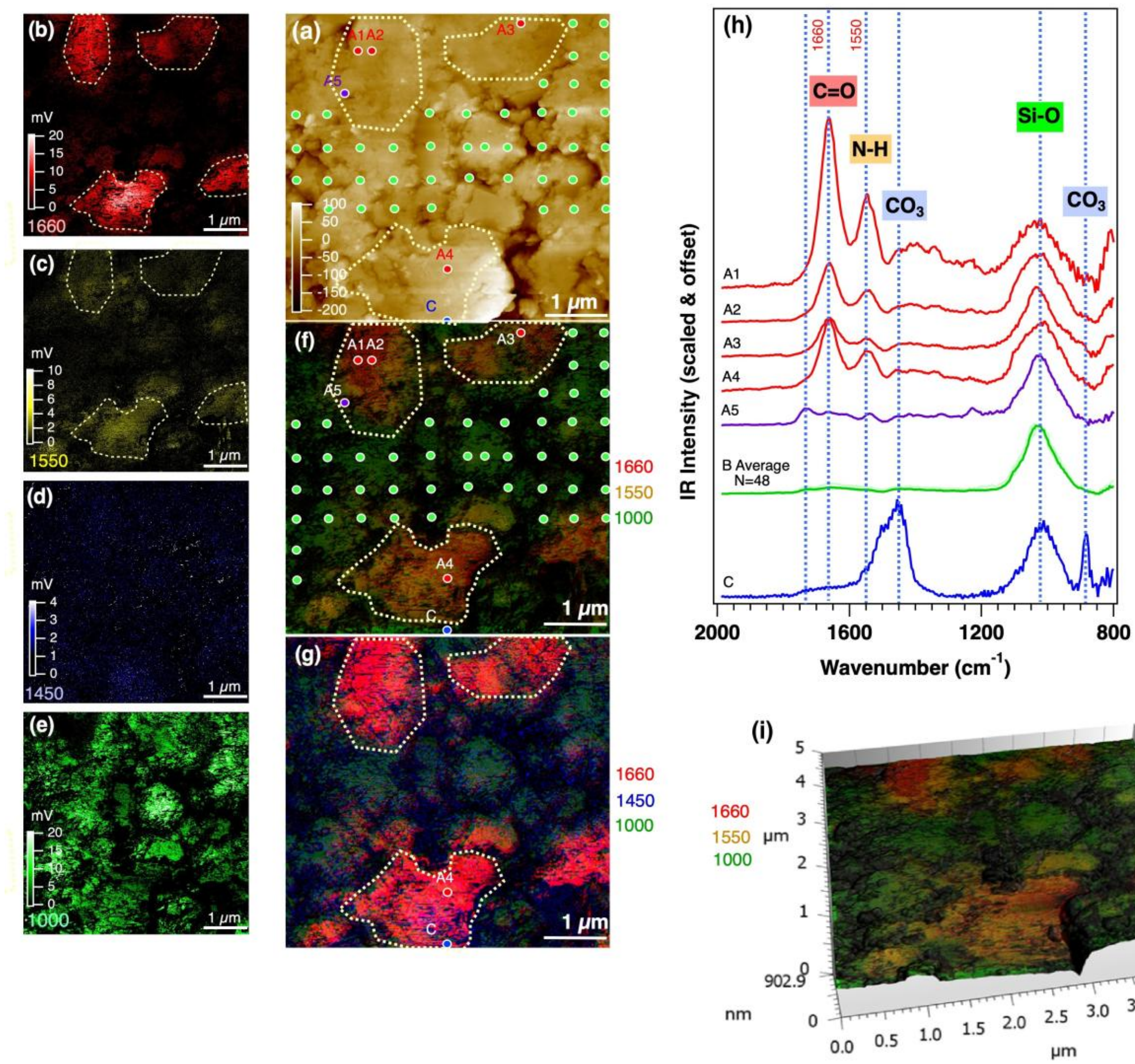

**Figure 3.** AFM-IR tapping IR maps using the APE laser (2000-700 $cm^{-1}$) of the 7-G2-1 region with (a) AFM image from particle C0052 at spectral bands of (b) 1660 $cm^{-1}$ and (c) 1550 $cm^{-1}$ (C=O from organic matter and C=O from amide group and $H_2O$); (d) 1450 $cm^{-1}$ (carbonate bending mode or ammonium); and (e) 1000 $cm^{-1}$ (phyllosilicates); and combined (f) RGY and (g) RGB maps (R = 1660 $cm^{-1}$; G = 1000 $cm^{-1}$, Y = 1550 $cm^{-1}$, B = 1450 $cm^{-1}$). The organic-rich region is highlighted by white dashed lines. (h) AFM-IR contact spectra associated with different positions located in the AFM topographic map. AFM-IR spectra of A1-A4 show double peaks at ~1660 $cm^{-1}$ and ~1550 $cm^{-1}$ possibly related to C=O stretching and -NH bending vibration mode of amide group (-C(=O)-NH-), respectively. The vertical blue dashed lines highlight the characteristic band positions of C=O and N-H from the amide group at 1660 and 1550 $cm^{-1}$ (black), $NH_4^+$ at 1450 $cm^{-1}$ (yellow), phyllosilicates at 1000 $cm^{-1}$ (green), and carbonates ($CO_3$) at 1440 and 880 $cm^{-1}$ (blue). (i) composite 3-D view of the RGY image derived from the three maps: 1660 $cm^{-1}$, 1000 $cm^{-1}$, and 1550 $cm^{-1}$, respectively, showing the areas containing the amides.

The 7-G2-1 region was also analyzed in contact mode using the Firefly laser (3800−2700 $cm^{-1}$ range) across areas with and without amides (Figure 4). AFM-IR images were obtained at 2930 $cm^{-1}$ (red), 2960 $cm^{-1}$ (pink), 3240 $cm^{-1}$ (blue) and 3680 $cm^{-1}$ (green) corresponding respectively to 2930 $cm^{-1}$ and 2960 $cm^{-1}$, the asymmetric stretching modes of $-CH_2$ and $-CH_3$ groups in organic matter (OM), 3240 $cm^{-1}$, possible −N-H stretching mode and 3680 $cm^{-1}$, −OH stretching mode from phyllosilicate. This analysis revealed a significant contribution from −OH stretching at 3680 $cm^{-1}$ and strong absorption at 2930 $cm^{-1}$ and 2960 $cm^{-1}$, highlighting aliphatic asymmetric stretching modes in OM. The aliphatic signatures were more pronounced in the region associated with amides (region A in Figure 4) compared to those associated with phyllosilicates (region B in Figure 4). Peaks in the range of 3358 $cm^{-1}$ to 3240 $cm^{-1}$ were detected in the region associated with amides, possibly corresponding to the characteristic N-H stretching mode of amide (Pelton and McLean, 2000). However, it remains unclear whether these peaks definitively correspond to N-H stretching of amide due to the low signal-to-noise ratio and potential physical damage to the AFM tips during IR mapping in contact mode.

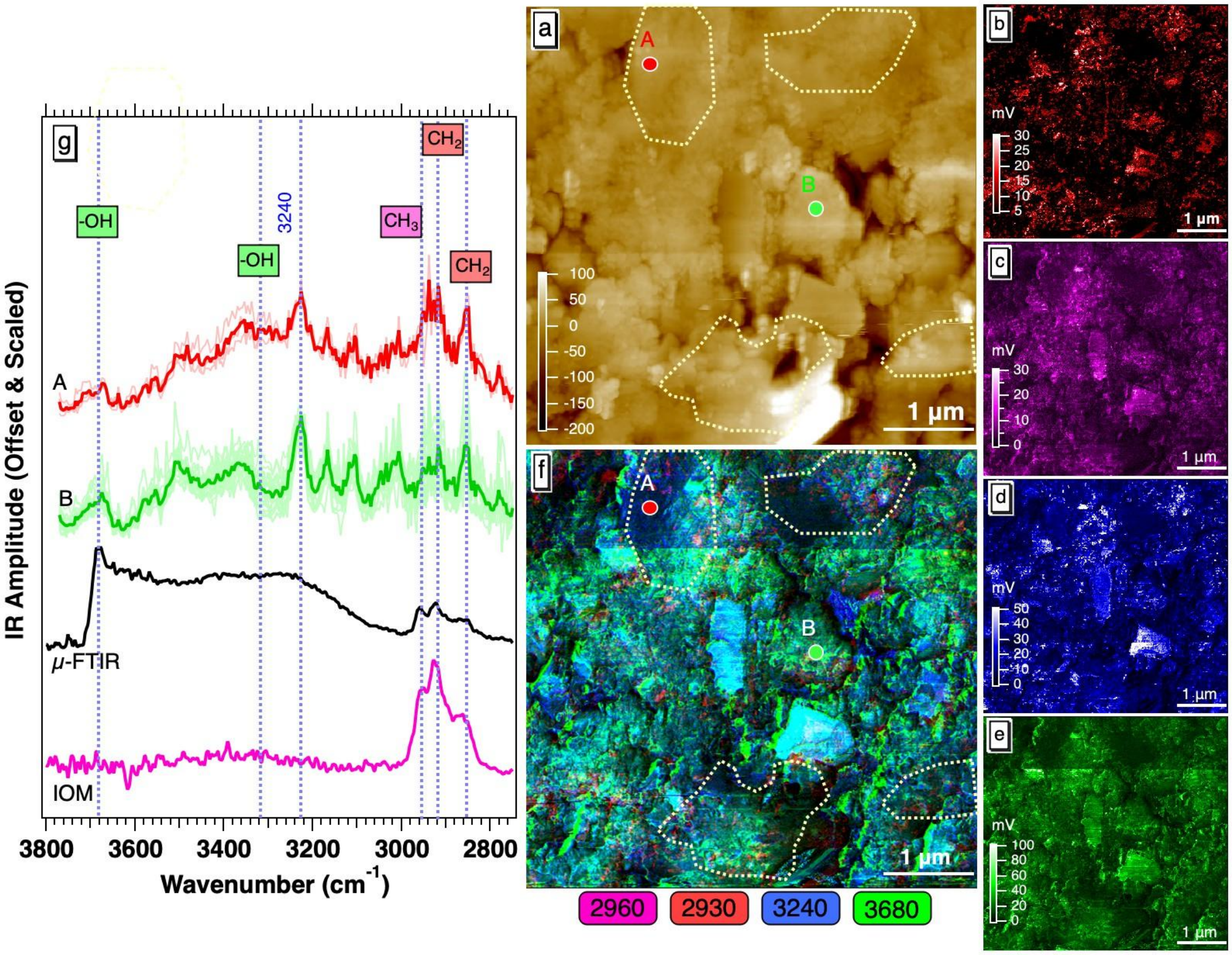

**Figure 4.** AFM-IR tapping mode maps acquired with the FF laser (4000−2700 $cm^{-1}$) of the region 7-G2-1 also analyzed with the APE laser (Figure 3). (a) Topographical image. (b) Aliphatic $CH_2$ asymmetric stretching at 2930 $cm^{-1}$ (red). (c) Aliphatic $CH_3$ asymmetric stretching at 2960 $cm^{-1}$ (pink). (d) Possible N-H stretching at 3240 $cm^{-1}$. (e) −OH stretching at 3680 $cm^{-1}$. (F) RGB (Red, Green, Blue) composite image of the three maps: 2930 $cm^{-1}$, 3680 $cm^{-1}$ and 3240 $cm^{-1}$, respectively. (g) AFM-IR contact spectra associated with regions A (red) and B (green) located in the AFM topographic map, compared to the μ-FTIR spectra of C0052 grain (black) and the insoluble organic matter (IOM) extracted from the C0057 grain (pink, Quirico et al., 2024). The vertical blue dashed lines highlight the characteristic band positions of -OH from phyllosilicates, aliphatic signatures ($CH_2$−$CH_3$) from organic matter, and N-H stretching mode at 3240 $cm^{-1}$ from amide compounds.

### 3.4. NanoSIMS

High spatial resolution NanoSIMS mapping (5 × 5 µm² of 256 × 256 pixels) was conducted on the same area where the amides were detected by AFM-IR (Figures 3−4), as well as in surrounding areas to further investigate the presence of these compounds (Figures S5−S6). In the first session, we collected $^{12}C^-$, $^{13}C^-$, $^{12}C^{14}N^-$, and $^{12}C^{15}N^-$ ions to determine their respective $\delta^{13}C$ and $\delta^{15}N$ values. Using AFM-IR tapping mode map at 1660 $cm^{-1}$ and NanoSIMS accumulated images of $^{12}C^{14}N^-$ ions as reference images, we defined ROIs on the corresponding NanoSIMS images to select only pixels from the amide areas (Figure 5, Table 1). Given that our AFM-IR investigations were focused solely on the 7-G2-1 region, we inferred that the other $^{12}C^{14}N^-$-rich compounds detected by NanoSIMS in the surrounding areas, exhibiting similar $^{12}C^-$ and $^{12}C^{14}N^-$ counts per pixel and comparable size (Table 1), likely correspond to the same amides identified by AFM-IR (Figures S5−S6).

The C isotopic compositions of the amides are within the range of terrestrial values, with $\delta^{13}C$ values ranging from ~ −70 ‰ to 0 ‰, with a weighted mean of −26 ± 63 ‰ (2σ, $n$ = 6). In contrast, their N isotopic compositions are highly depleted in $^{15}N$, with $\delta^{15}N$ values ranging from ~ −300 ‰ to −100 ‰, yielding a weighted mean of −213 ± 103 ‰ (2σ, $n$ = 8). Their calculated wt% N/C ratios indicate that these organic compounds are enriched in N compared to their surrounding matrix, with N/C ratios between 0.10 and 0.14, averaging 0.13 ± 0.03 (2σ, $n$ = 6, Table 1).

In the second session, we collected $^{1}H^-$ and D ions to measure the δD values of the same ROIs defined during the first session. Following the C and N analysis, and due to sample consumption, we relied on secondary electron (SE), $^{12}C^{14}N^-$, and $^{1}H^-$ images to identify the remaining N-H organic compounds and establish ROIs for H isotope analysis (Figures 5 and S6). Although we systematically compared the last cycle of $^{12}C^{14}N^-$ with the first cycle of $^{1}H^-$ images to locate the remaining amide grains based on their shape and size, we cannot exclude the possibility that some of these grains may have been completely consumed in few analyses after C and N measurements. Nevertheless, the H isotopic compositions of the amides are not anomalous in D compared to terrestrial values, with δD values ranging from ~ −100 to 400 ‰, with a weighted mean of 194 ± 368 ‰ (2σ, $n$ = 7, Table 1).

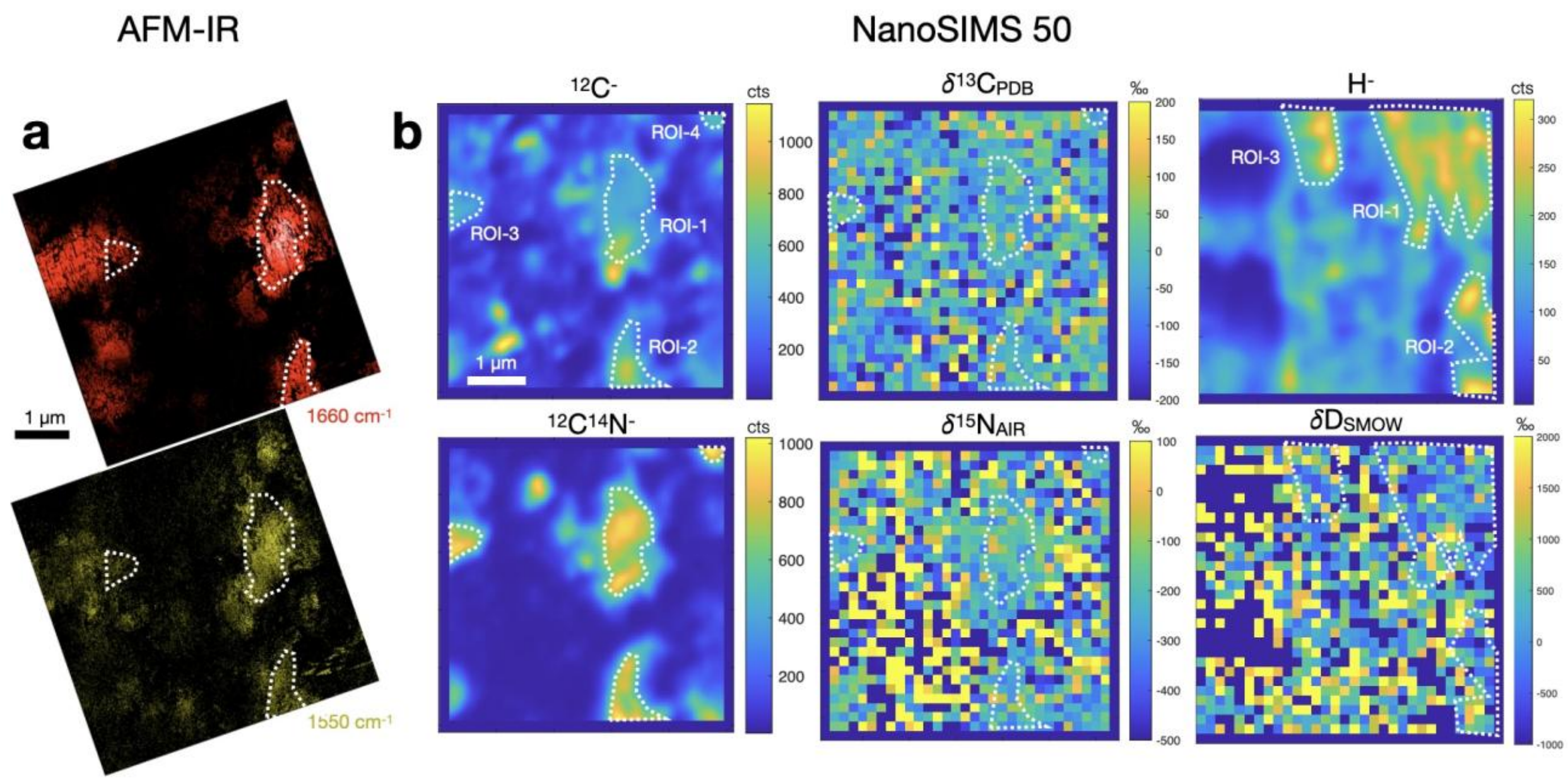

**Figure 5.** Combined AFM-IR and NanoSIMS high resolution spatial analyses of the amides. (a) AFM image of the 7-G2 region from particle C0052 at spectral bands of 1660 $cm^{-1}$ (C=O from amide I band) and 1550 $cm^{-1}$ (−NH bending vibration mode from amide II band). (b) Corresponding NanoSIMS accumulated images (5 × 5 $\mu m^2$) of $^{12}C^-$, $^{12}C^{14}N^-$ and $^1H^-$ ions, $\delta^{13}C$, $\delta^{15}N$ and $\delta D$. The amides regions detected by AFM-IR are outlined by white dashed lines. The shapes of the ROIs in the $^{12}C^{14}N^-$ and $H^-$ NanoSIMS images differ due to sample consumption (Figure S6).

**Table 1 -** C, N and H isotopic composition of the amides determined by NanoSIMS.

| Sample | ROI size ($\mu m^2$) | $^{12}C$ (cts/px) | $\delta^{13}C$ (‰) | 2σ | $^{12}C^{14}N$ (cts/px) | $\delta^{15}N$ (‰) | 2σ | N/C (wt%) | 2σ | ROI size ($\mu m^2$) | $^1H$ (cts/px) | $\delta D$ (‰) | 2σ |
|---|---|---|---|---|---|---|---|---|---|---|---|---|---|
| ROI-1 | 1.12 | 527 | 0 | 29 | 748 | -198 | 38 | 0.12 | 0.07 | 2.83 | 219 | 190 | 245 |
| ROI-2 | 0.49 | 602 | -7 | 33 | 716 | -218 | 50 | 0.10 | 0.06 | 0.44 | 236 | 130 | 363 |
| ROI-3 | 0.2 | 453 | -3 | 47 | 758 | -227 | 70 | 0.14 | 0.08 | 0.55 | 238 | 390 | 319 |
| ROI-4 | 0.08 | 530 | -69 | 66 | 860 | -215 | 100 | 0.14 | 0.08 | n.a. | | | |
| ROI-5 | 1.03 | – | – | – | 1131 | -123 | 34 | | | 1.27 | 146 | 121 | 308 |
| ROI-6 | 0.25 | – | – | – | 1198 | -147 | 51 | | | 0.54 | 182 | 429 | 344 |
| ROI-7 | 1.48 | 535 | -34 | 28 | 820 | -237 | 35 | 0.13 | 0.08 | 1.90 | 198 | -121 | 277 |
| ROI-8 | 0.92 | 260 | -6 | 35 | 396 | -307 | 52 | 0.13 | 0.08 | 0.48 | 466 | 159 | 291 |
| **Weighted mean** | | **485** | **-26** | | **828** | **-213** | | **0.13** | | | **214** | **194** | |
| *2σ* | | *239* | 63 | | 502 | 103 | | *0.03* | | | *209* | *368* | |

Note. ‘‘–” denotes discarded values; ‘‘n.a.” = not analyzed.

## 4. Discussion

### 4.1. Detection and identification of NH-bearing compounds in Ryugu and CI chondrites

A faint absorption band at 3.06 µm is present on average spectra of Ryugu grains and on a few individual particles (Yada et al., 2022; Pilorget et al., 2022). Jiang et al. (2024), analyzed numerous Ryugu grains at JAXA curation using MicrOmega, identifying only rare regions, a few hundred µm in size, with a prominent ~3.06 µm absorption feature. This spectral signature has previously been attributed to $NH_4^+$ phyllosilicates, $NH_4^+$ hydrated salts, $NH_4^+$ phosphorous-rich grains, and/or NH-bearing organic molecules (Yada et al., 2022; Pilorget et al., 2022, 2024). Since then, hot water extractions performed on Ryugu particles A0106 and C0107 have revealed the presence of amines ($R\text{-}NH_2$), hypothesized to be mainly in the form of $NH_4^+$ salts (Naraoka et al., 2023; Yoshimura et al., 2023; Takano et al., 2024), and $NH_4^+$ was detected in A0106 (Yoshimura et al., 2023) although at a much lower concentration than in samples of Bennu (Glavin et al., 2025) and Orgueil (Laize-Générat et al., 2024).

The faint absorption band around 3.06 µm, attributed to NH-rich compounds, observed in spectral average of several Ryugu grains within the curation facility (Yada et al., 2022), is not observed here in mm-scale spectra of the bulk C0050 and C0052 particles (Figure 1). This is most probably because the width of the 2.7-µm absorption band extends up to more than 3.1 µm because of the adsorption of water molecules, thus masking any faint absorption potentially present. However, absorption around ~3250 $cm^{-1}$ was observed locally at µm-scale, on crushed fragments of these grains (Figure 2).

Here, we report the detection of amides (-C(=O)-NH-bearing organic compounds) at sub-micrometer-scale by AFM-IR (Figures 3−4), in locations of Ryugu grains where the N-H absorption band was possibly detected at micrometer-scale by µ-FTIR (Figure 2). In total, we analyzed ~7380 $\mu m^2$ of surface area across the C0050 and C0052 Ryugu particles using AFM-IR and detected amide-related organic compounds in only ~35 $\mu m^2$ (Tables S1−S2), representing ~1 vol.% of the total analyzed surface. This result suggests that amides are relatively rare in our Ryugu samples and are heterogeneously distributed at a micrometer scale. The amide N-H stretching vibration gives the so-called "amide A" band between 3310 $cm^{-1}$ and 3270 $cm^{-1}$ (3.02−3.06 µm) which is usually part of a Fermi resonance with a second component (so-called "amide B" band) absorbing weakly between 3100 $cm^{-1}$ and 3030 $cm^{-1}$ (3.22−3.30 µm) (Barth, 2007). Therefore, amides are likely contributors to the ~3.06 µm absorption feature observed on Ryugu particles, together with $NH_4^+$ and amines.

In meteorites, amide-related organic compounds have been previously reported in aqueously altered carbonaceous chondrites. The quantities of amino acids and amines extracted from carbonaceous chondrites generally increase by a factor of two, and sometimes much more (up to about 10), when an acid hydrolysis is performed following, or instead of, a water extraction. Therefore, a significant fraction of these molecules is initially in the form of acid-labile precursors, such as peptide-like structures and amides (Cronin, 1976a; Pizzarello and Holmes, 2009; Burton et al., 2014; Aponte et al., 2014; Kebukawa et al., 2019a; Glavin et al., 2020). Such increases of amino acid yields after acid hydrolysis were also observed on lunar regolith samples, and potentially attributed to precursors contained in meteorites, micrometeorites, or inter-planetary dust particles (Elsila et al., 2016). Analyses performed so far on the A0106 and C0107 Ryugu grains revealed that most amino acids were extracted in hot water, without acid hydrolysis. A notable exception is γ-amino-n-butyric acid whose abundance increased by a factor of more than 15 after acid hydrolysis, indicating that it is present in the form of precursors possibly including the amides γ-lactams (Parker et al., 2023).

Several studies have investigated acid-labile amino acid precursors in carbonaceous chondrites, particularly in the CM2 Murchison, highlighting diverse molecular origins. Acid hydrolysis has been shown to convert amides, including aliphatic amides in Murchison and polymer amides in both Murchison and CV3 Allende, into amino acids (Cronin, 1976a, b; Cooper and Cronin, 1995; McGeoch and McGeoch, 2015). Meierhenrich et al. (2004) reported higher concentrations of diamino acids in Murchison after acid hydrolysis compared to water extraction, suggesting these compounds originally existed as higher-molecular-mass precursors. Shimoyama and Ogasawara (2002) detected the presence of dipeptide (glycyl-glycine) in Murchison and CM2 Yamato-791198, while Lange et al. (2020) found that most amino acids in acid-hydrolyzed meteorite extracts originate from non-peptide precursors, though they confirmed the presence of peptide sequences in Murchison. In Ryugu samples, numerous N-bearing organic molecules, such as N-heterocycles, amines, hydroxy compounds, and N-heterocyclic indoles, have been identified via hot water extraction (Naraoka et al., 2023; Oba et al., 2023; Takano et al., 2024). However, amides or polypeptides acid-labile precursors have yet to be recovered from Ryugu samples.

FTIR analysis of the acidic residues from Ryugu particles, however, revealed two distinct phases: one resembling the insoluble organic matter (IOM) typical of unheated carbonaceous chondrites type 1 and 2 (Kebukawa et al., 2011; Orthous-Daunay et al., 2013; Quirico et al., 2024), and the other representing an "outlier phase" (Kebukawa et al., 2024). Notably, a peak at 1660 $cm^{-1}$, attributed to C=O stretching in amide compounds (R-(C=O)-N-H-R') or unsaturated ketones/aldehydes, was consistently observed in residues C0107, A0106, and C0002 (Kebukawa et al., 2024) (Figure 6). Furthermore, two absorption bands at 3350 $cm^{-1}$ and 3180 $cm^{-1}$ were detected in the "outlier phase" of residues A0106 and C0107, likely corresponding to N-H stretching. Interestingly, the peak at 1550 $cm^{-1}$, commonly associated with N-H bending in amide compounds, was absent in all spectra of these residues (Kebukawa et al., 2024). This outlier phase was exclusively observed in the acid residues of C0107, A0106, and C0002 which was extracted using HF/HCl demineralization and following solvent extraction protocols, as described by Yabuta et al. (2023) and Naraoka et al. (2023). In contrast, this outlier phase was not observed in acid residues prepared by Quirico et al. (2023), including A0106, which were obtained using a different HF/HCl protocol (Orthous-Daunay et al., 2010). This discrepancy could be attributed to sample heterogeneities (Kebukawa et al., 2024), especially given the small sample amount used by Quirico et al. (2024) compared to Yabuta et al. (2023). However, the possibility of contamination cannot be ruled out. Kebukawa et al. (2024) demonstrated that the IR spectra of very simple molecules closely match the spectrum of this outlier phase. This finding is somewhat unexpected, given the broad diversity of organic species in soluble organic matter (SOM) (Schmitt-Kopplin et al., 2023) and the observed decrease in abundance with increasing carbon chain length (Kebukawa et al., 2024).

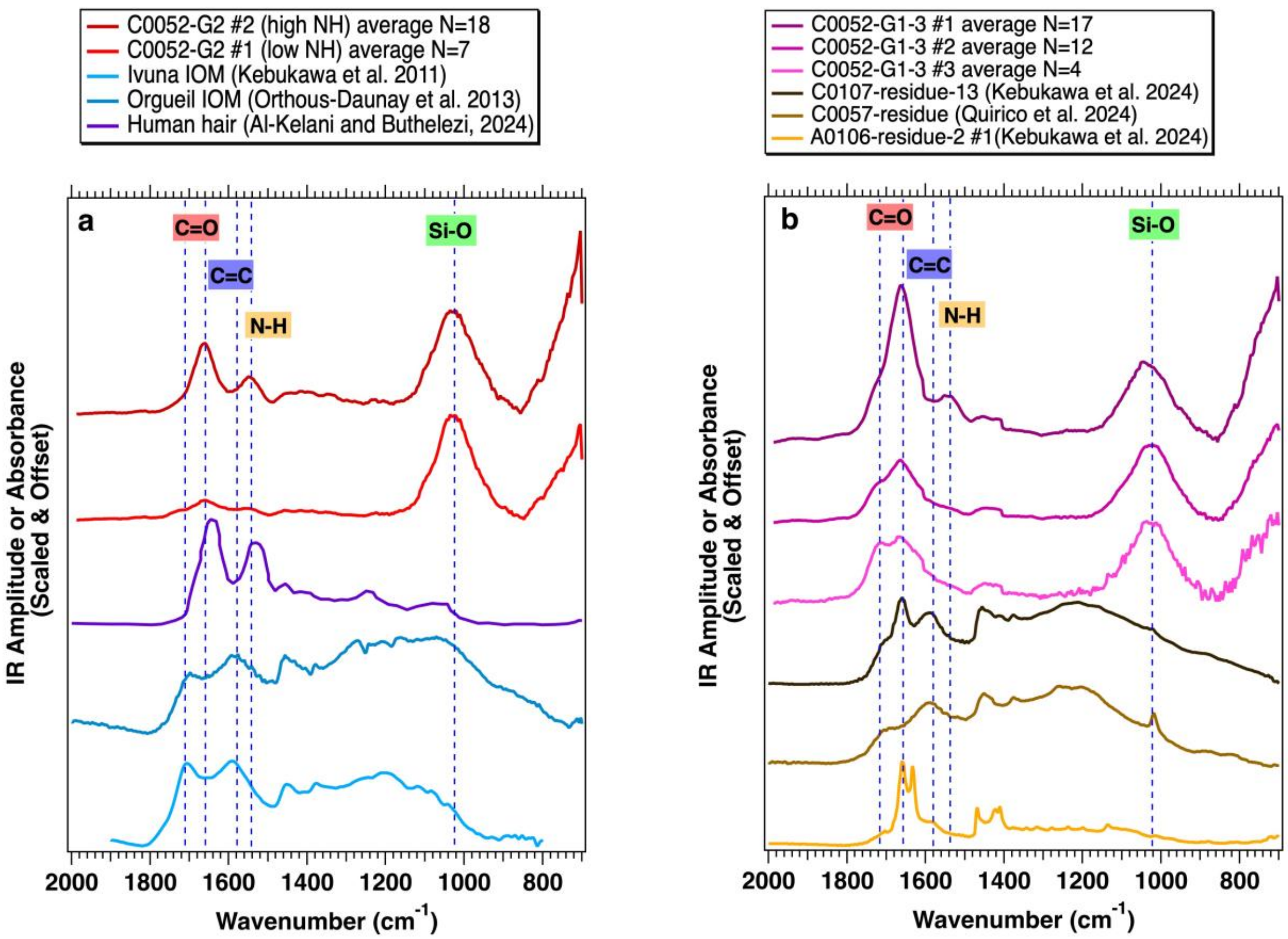

**Figure 6.** AFM-IR spectra of Ryugu grains C0052: (a) 7-G2 and (b) 7-G3 showing key vibrational bands for C=O, C=C, N-H, and Si-O (indicated by blue dashed lines). For comparison, the IR spectra of Ivuna IOM (Kebukawa et al., 2011), Orgueil IOM (Orthous-Daunay et al., 2013), and human hair (Al-Kelani and Buthelezi, 2024) are included, along with acid residues from Ryugu samples C0057 (Quirico et al., 2024), A0106, and C0107 (Kebukawa et al., 2024). Spectra are scaled and offset for clarity.

### 4.2 Potential source of contaminations

Given that the amides were detected solely in a single preparation of C0052 (Figure S3), one possibility is that this phase resulted from terrestrial contamination either before shipment to our laboratory or during sample preparation and/or IR measurements. Concerning the former, Kebukawa et al. (2024) discussed possible contamination sources in Ryugu samples, including artificial OM particles inside sample containers at the JAXA Curation Center (Yada et al., 2014; Uesugi et al., 2014) or explosives contaminants introduced during FTIR measurements (Takano et al., 2020; Ito et al., 2021). However, a comparison of these potential contaminants with the spectra from the C0052 grain reveals no spectral match with the amide phases identified in C0052 through AFM-IR analysis.

Another possibility of contamination is during sample preparation and measurements in our laboratory. Contamination from human skin or hair, for example, could have been introduced during sample handling. Kebukawa et al. (2024) provided an IR spectrum extracted from human skins treated with HF/HCl, and FTIR spectra of untreated human skin or hair are available in the literature (Wang et al., 2012; Al-Kelani and Buthelezi, 2024) (Figure 6a). The two peaks at 1645 $cm^{-1}$ and 1530 $cm^{-1}$ from the human hair are fairly consistent with the amide phase in Ryugu C0052 (1660 $cm^{-1}$ and 1550 $cm^{-1}$), which could suggest a contamination origin. However, the N isotopic composition of the amides in Ryugu C0052 is significantly depleted in $^{15}N$ ($\delta^{15}N \simeq -200$ ‰, Table 1) compared to human hair samples, which have a bulk $\delta^{15}N$ of ~10 ‰ (Petzke et al., 2005). These isotopic differences strongly suggest that the amide phase in Ryugu C0052 grains is indigenous to the samples, rather than the result of a laboratory contamination.

## 4.3 Possible mechanisms for the formation of the amides in Ryugu samples

### *4.3.1 Hydrothermal condition on the Ryugu's parent planetesimal?*

Abiotic formation of amides is feasible under low-temperature hydrothermal conditions (≤ 300 °C) through condensation of carboxylic acids with $NH_3/NH_4^+$ or amines, followed by dehydration (Figure 7) (Rushdi and Simoneit, 2004; Lange et al., 2020; Fu et al., 2020; Kebukawa et al., 2024; Serra et al., 2024). Indeed, carboxylic acids, $NH_4^+$, and aliphatic amines have been detected in Ryugu samples and carbonaceous chondrites, particularly in hot water extracts (Sephton, 2002; Naraoka et al., 2023; Yoshimura et al., 2023; Schmitt-Kopplin et al., 2023; Takano et al., 2024). The presence of these compounds, along with evidence of extensive aqueous alteration on Ryugu's parent asteroid, supports the hypothesis that amides in Ryugu could form via hydrothermal alteration (Kebukawa et al., 2024).

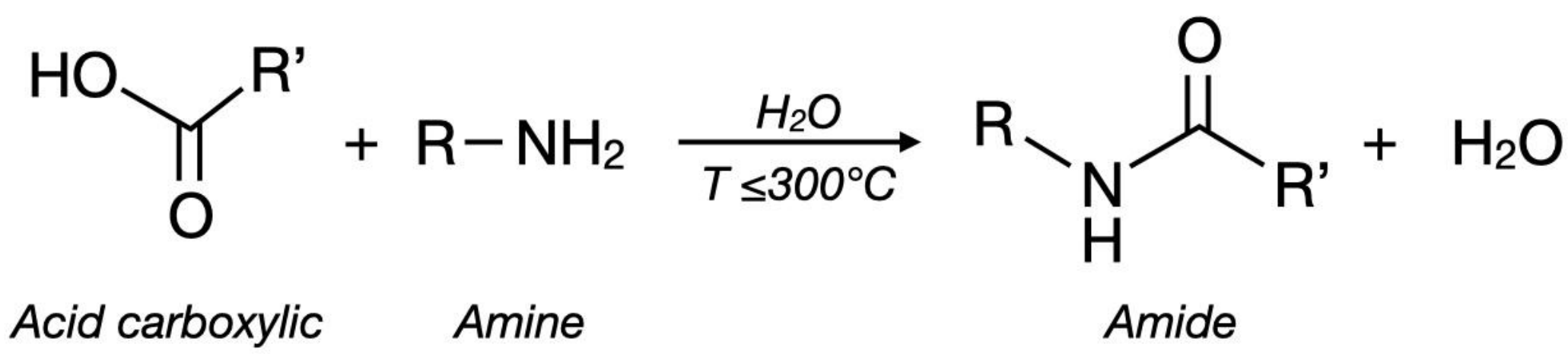

**Figure 7.** Hydrothermal formation of amides though condensation of amines and carboxylic acids (adapted from Fu et al., 2020). R and R′ represent generic alkyl or aromatic groups.

Assuming the amides in Ryugu formed through the condensation of carboxylic acids and amines/$NH_4^+$, their N, C, and H isotopic signatures should approximately align with those of their expected precursor molecules (Marlier et al., 1999). Takano et al. (2024) reported the C and N isotopic compositions of water-soluble organic molecules (WSOM) extracted from Ryugu samples, including carboxylic acids and amines. Their $\delta^{13}C$ values were consistent within error ($\delta^{13}C \simeq -30$ ‰ to $-20$ ‰) with the amide phase observed in this study (Figure 8a). However, their $\delta^{15}N$ values ($\delta^{15}N \simeq 0$ ‰ to 60 ‰) were significantly enriched in $^{15}N$ when compared to the extremely depleted $\delta^{15}N$ value of the amide phase in Ryugu ($\delta^{15}N \simeq -200$ ‰, Figure 8a, Table 1).

Additionally, Becker and Epstein (1982) measured δD values from the SOM in various carbonaceous chondrites (δD ≃ −50 ‰ to 490 ‰), showing ranges comparable to those of Ryugu's amide phase (δD ≃ 200 ‰, Figure 8b and Table 1). Despite similar $\delta^{13}C$ and $\delta D$ values, the $\delta^{15}N$ of the amides in Ryugu are strikingly depleted in $^{15}N$ compared to the WSOM and SOM from both Ryugu and carbonaceous chondrites (Figure 8). This isotopic distinction indicates that the amide phase observed in this study is distinct from other N-bearing phases, potentially reflecting specific formation pathways under distinct environmental conditions.

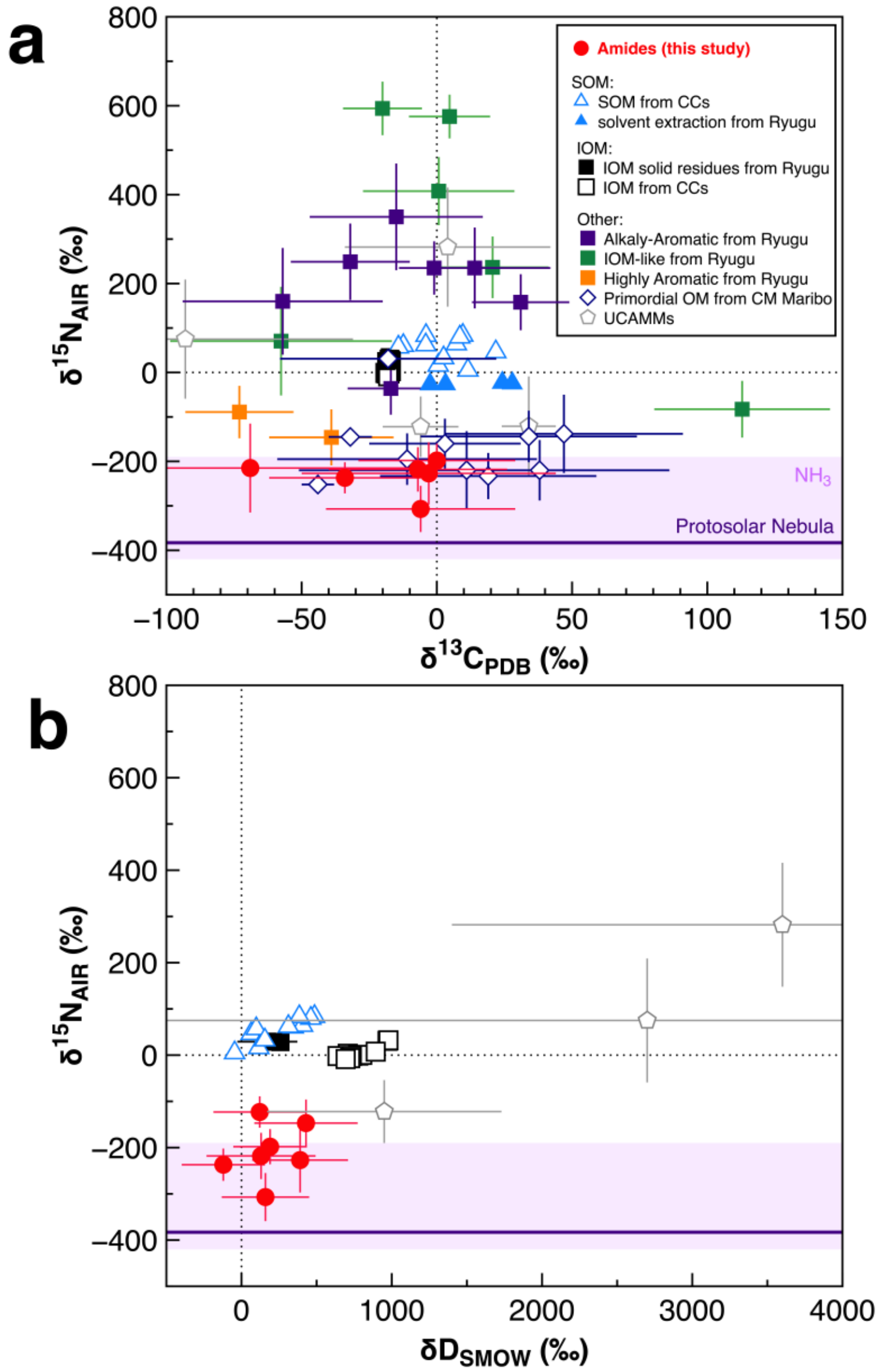

**Figure 8.** C, N and H isotopic compositions of the amides (red circles) detected in Ryugu's particle C0052 compared to literature data. (a) Plots of δ$^{15}$N *vs.* δ$^{13}$C and (b) δ$^{15}$N *vs.* δD for various classes of organic matter from Ryugu samples and carbonaceous chondrites, and ultra-carbonaceous Antarctic micrometeorites (UCAMMs, Rojas et al., 2024). The amides are compared to soluble organic matter (SOM), insoluble organic matter (IOM), and other C-rich organic grains from Ryugu (Takano et al., 2024; De Gregorio et al., 2024) and carbonaceous chondrites (Becker and Epstein, 1982; Alexander et al., 2007; Vollmer et al., 2020b). Black dashed lines correspond to terrestrial references, the solid purple line correspond to the $\delta^{15}$N value of the protosolar nebula (Marty et al., 2011), and the colored pink region indicate the range of $\delta^{15}$N values of $NH_3$ in cold molecular clouds (Gerin et al., 2009; Lis et al., 2010). Uncertainties are 2σ.

*4.3.2 Irradiation of $^{15}$N depleted N-bearing ice?*

Comparable depletions in $^{15}$N as those found in Ryugu's amides have been observed in "coldspots" within OM-like C-rich grains or nanoglobules in Ryugu samples (Nguyen et al., 2023; Yabuta et al., 2023; Nittler et al., 2024; Stroud et al., 2024; De Gregorio et al., 2024), CM Maribo (Vollmer et al., 2020b), CO DOM 08006 (Nittler et al., 2018), CR chondrites (Floss and Stadermann, 2009; Floss et al., 2014; Vollmer et al., 2020a), and ultra-carbonaceous Antarctic micrometeorites (UCAMMs; Rojas et al., 2024) (Figure 8). However, few studies have examined both the chemical and isotopic properties of these $^{15}$N-depleted organic compounds in details. De Gregorio et al. (2024) reported two Ryugu grains in A0108-3, both depleted in $^{15}$N and $^{13}$C, with highly aromatic functional groups. In CM Maribo, Vollmer et al., (2020b) reported primordial $^{15}$N-depleted organic material ($\delta^{15}$N $\simeq -250$ ‰ to 30 ‰, Figure 8a) dominated by aromatic and ketone/carbonyl (C=O) bonds, as well as C=N/C≡N bonds. They suggested that this material may share an isotopic composition with HCN molecules from the ISM. In addition, Rojas et al., (2024) identified extremely light N bulk isotopic signatures in two UCAMMs rich in uncharacterized OM ($\delta^{15}$N $\simeq -120$ ‰, Figure 8a), proposing that these grains provide evidence of galactic cosmic rays (GCR) irradiation of primordial $^{15}$N-depleted $N_2$ ice on surface of cold objects from the outer regions of the Solar System.

The $^{15}$N-depleted signature of the amides in our Ryugu samples is consistent with their formation from a $^{15}$N-depleted reservoir. Amide-related components (e.g., formamide) can form through the irradiation of N-bearing ices ($N_2$ and $NH_3$) exposed to ultraviolet (UV) photons, energetic electrons or ions at low temperatures (<50 K) (Briggs et al., 1992; Gerakines et al., 2004; Jones et al., 2011; Kaňuchová et al., 2016). The protosolar nebula ($\delta^{15}$N $\simeq -380$ ‰, Marty et al., 2011) and $NH_3$ in cold molecular clouds ($\delta^{15}$N between $-420$ ‰ and $-190$ ‰, (Gerin et al., 2009; Lis et al., 2010) exhibit lower or similar $\delta^{15}$N compared to the N-H organic compounds in Ryugu. The isotopic fractionation of N during the formation of organic materials through ice irradiation remains poorly understood (Almayrac et al., 2022). However, experiments with UV and heavy ions irradiations of NH-bearing ice indicate little to no $^{15}$N and D enrichment in the synthetized OM compared to the initial ices (Augé et al., 2019; Sugahara et al., 2019).

Therefore, a plausible scenario for the formation of $^{15}$N-depleted amides in our Ryugu samples could involve the UV or GCR irradiations of $^{15}$N-depleted $N_2$ or $NH_3$-bearing ices either *(i)* at the surface of the Ryugu's parent planetesimal in the outer Solar System, or *(ii)* on mantle of interstellar dust grains in the ISM before their accretion on the planetesimal. The former is consistent with other evidence indicating that Ryugu's parent asteroid (and CI chondrites) likely formed beyond the $H_2O$ and $CO_2$ snow lines (i.e., ≥ 3−15 AU, Desch et al., 2018; E. Nakamura et al., 2022; Hopp et al., 2022; Ito et al., 2022; Piani et al., 2023; Brunetto et al., 2023; Matsumoto et al., 2024; Maurel et al., 2024). However, the penetration of UV photons into chondritic material is weak (~100 nm), making their contribution highly unlikely. The penetration of solar ions and GCR have a larger range, from nm to ~1 m. Over 4.55 billion years, the accumulated dose can be as large as 10 eV/C atom. This raises the question about the exposure time of the Ryugu material at the surface of the parent body. Given that Ryugu formed from the recent collision of a large asteroid (Potiszil et al., 2023), it is clear that the exposure time ranges from zero to a few eV/C atoms. Consequently, no firm conclusion can be drawn. The exposure time at the surface of grains in the local ISM, or possibly in the outer regions of the protosolar disk, is around a few million years.

Experimental simulations show that the effect of UV photons and swift heavy ions on $CH_3OH$:$NH_3$ mixtures are almost similar (Muñoz Caro et al., 2014). These experiments have identified amide functions in the molecular by-products, along with ions like $OCN^-$ and $NH_4^+$. Furthermore, irradiation-induced chemistry is observed at doses as low as 1 eV/C atom. Several studies also show that $N_2$ ice is radiolyzed under the effect of ion irradiation, leading to the formation of $NH_3$ and several N-rich molecular by-products (Palumbo et al., 2004; Augé et al., 2016; Rocha et al., 2020). It is therefore plausible that amide functions are formed by ion irradiation or UV photons in the ISM or the protosolar disk. However, due to the limited number of constraints, no definitive conclusion can be drawn.

## 5. Conclusions

In this study, we report the multi-scale (millimeter to nanometer) characterization of amides in two Ryugu's particles (C0050 and C0052) from Chamber C. High-resolution AFM-IR imaging confirmed the presence of rare (~1 vol.%), micrometer-sized amides (-C(=O)-NH-bearing organic compounds), unevenly distributed in Ryugu's particles. These amides are potential contributors, together with $NH_4^+$ and amines, to the faint ~3.06 µm absorption band observed at millimeter-scale on the global Ryugu grain collection (Yada et al., 2022, Pilorget et al., 2022). Combined AFM-IR and NanoSIMS analysis revealed that these amides have C and H isotopic compositions compatible with OM found in carbonaceous chondrites ($\delta^{13}C \simeq -20$ ‰ and $\delta D \simeq 200$ ‰), but they differ significantly in their N isotopic composition, being highly $^{15}N$-depleted ($\delta^{15}N = -200$ ‰). Although no amides have been recovered in Ryugu samples by chemical extraction so far, the $^{15}N$-depleted signature of the amides supports their indigenous origin in our samples.

The amides may have formed through aqueous alteration on the Ryugu's parent planetesimal or irradiation of $^{15}N$-depleted reservoir(s). Hydrothermal conditions could have led to amide formation from carboxylic acids and amine/$NH_4^+$ precursors on the Ryugu's parent planetesimal, although isotopic analysis suggests that water-soluble organic molecules in Ryugu are enriched in $^{15}N$ relative to the amides. Alternatively, the amides could have formed through irradiation of $^{15}N$-depleted ice (such as $N_2$ or $NH_3$) by UV photons or galactic cosmic rays in the outer Solar System or ISM.

C-type asteroids and comets may have delivered light elements and volatile organics to the inner Solar System, including early Earth (Chyba and Sagan, 1992; Martins and Pasek, 2024). Amides delivered on the early Earth by objects such as Ryugu, may have played a role in prebiotic chemistry.

**Acknowledgments**

The authors are very grateful to JAXA/ISAS Astromaterial Science Research Group for allowing the allocation of the two Ryugu's particles C0050 and C0052 to this research consortium through the first international Ryugu sample AO. We also acknowledge Frédéric Charlot for the assistance during SEM/EDS analysis at CMTC (Grenoble, France). This work has received support from H2020 European Research Council (ERC) (SOLARYS ERC-CoG2017_771691) and the Centre National d'Etudes Spatiales (CNES) within the framework of the Hayabusa2 and MMX missions. LGV also acknowledges support from CNES, and Agence Nationale de la Recherche (ANR) CLASTS (ANR-22-CE49-0013) through post-doctoral grants. LGV, VTHP, EQ, LB, PB and OP acknowledge the funding from LabEx OSUG@2020 (Investissements d'avenir – ANR10 LABX56) and from the Centre National d'Etudes Spatiales (CNES), essential for maintaining the infrared measurements facilities at IPAG.

## Appendix

### Data availability

The raw reflectance and µ-FTIR spectra are available online through the SSHADE Solid Spectroscopy database infrastructure with the following references: Beck and Poch (2022) for reflectance spectra and Vacher and Phan (2022) for µ-FTIR spectra.